\documentclass[twocolumn,secnumarabic,amssymb, nobibnotes, aps, prb, superscriptaddress, longbibliography]{revtex4-2}
\usepackage{graphicx}
\usepackage{natbib}
\usepackage{xcolor}
\usepackage[colorlinks,allcolors=blue]{hyperref}
\usepackage[english]{babel}

\begin{document}



\title{Microcavity polaritons for topological photonics}

\author{Dmitry D. Solnyshkov}
\affiliation{Institut Pascal, PHOTON-N2, Universit\'e Clermont Auvergne, CNRS, SIGMA Clermont, F-63000 Clermont-Ferrand, France}
\affiliation{Institut Universitaire de France (IUF), F-75231 Paris, France}

\author{Guillaume~Malpuech}
\affiliation{Institut Pascal, PHOTON-N2, Universit\'e Clermont Auvergne, CNRS, SIGMA Clermont, F-63000 Clermont-Ferrand, France}

\author{Philippe~St-Jean}
\author{Sylvain~Ravets}
\author{Jacqueline~Bloch}
\affiliation{Universit\'e Paris-Saclay, CNRS, Centre de Nanosciences et de Nanotechnologies, 91120, Palaiseau, France}

\author{Alberto~Amo}
\affiliation{Univ. Lille, CNRS, UMR 8523 -- PhLAM -- Physique des Lasers Atomes et Mol\'ecules, F-59000 Lille, France}
\email{alberto.amo-garcia@univ-lille.fr} 



\begin{abstract}
Microcavity polaritons are light-matter quasiparticles that arise from the strong coupling between excitons and photons confined in a semiconductor microcavity. They typically operate at visible or near visible wavelengths. They combine the properties of confined electromagnetic fields, including a sizeable spin-orbit coupling, and the sensitivity to external magnetic fields and particle interactions inherited from their partly matter nature. 
These features make polaritons an excellent platform to study topological phases in photonics in one and two dimensional lattices, which band properties can be directly accessed using standard optical tools. In this review we describe the main properties of microcavity polaritons and the main observations in the field of topological photonics, which include, among others, lasing in topological edge states, the implementation of a polariton Chern insulator under an external magnetic field and the direct measurement of fundamental quantities such as the quantum geometric tensor and winding numbers in one- and two-dimensional lattices. Polariton interactions open exciting perspectives for the study of nonlinear topological phases.
\end{abstract}
\date{\today}%
\maketitle

\section{Introduction}
Since the early 2010's, the field of photonics has been at the forefront in the study of novel topological phases of matter~\cite{Haldane2008, Ozawa2019}. Research in photonic systems has played a key role in understanding the symmetries behind the emergence of topological effects~\cite{Wu2015b}, in the discovery of anomalous Floquet phases~\cite{Kitagawa2012, Maczewsky2017,Mukherjee2017a} and in the first observations of Weyl points~\cite{Lu2015} and Fermi arcs~\cite{JihoNoh2017} (see Refs.~\cite{Lu2014, Ozawa2019} for detailed review). It has inspired original concepts, such as topological lasing~\cite{Ota2020}, the topology of PT symmetric non-Hermitian systems~\cite{Ozdemir2019}, the study of quantum optics effects in topological landscapes~\cite{Barik2018, Blanco-Redondo2018} and the direct measurement of the geometry and topology of energy bands~\cite{Gianfrate2020}. Topological concepts have provided a very efficient route to engineer transmission channels in photonic chips with very low losses~\cite{Hafezi2011, Hafezi2013, Wu2015b}.

Probably, the greatest assets of photonic systems to address topological phenomena are (i) the possibility of engineering the photonic band structure in a very flexible way, in particular in two-dimensional lattices; (ii) the direct measurement of the photonic bands eigenstates and edge states in simple optical experiments; (iii) the use of active and nonlinear optical materials to explore novel topological phenomenology. Some of their disadvantages are the relatively weak sensitivity to external magnetic fields of photons at optical frequencies in photonic materials, which hinders the study of topological systems with broken time-reversal symmetry, the short photon lifetime of certain systems, and their weak nonlinearities.

A particularly suitable photonic platform to explore topological phenomena, which overcomes the above mentioned issues, is represented by microcavity polaritons~\cite{Carusotto2013}. The engineering of lattices of coupled polariton micropillars allows the implementation of lattice Hamiltonians well-described by tight-binding models~\cite{Schneider2017}, and the hybrid light-matter nature of polaritons makes them both sensitive to magnetic fields and subject to interactions. In this article, we review the main properties of the microcavity polariton systems and the main observations in the field of topological physics. 

\subsection*{Microcavity polaritons: hybrid light-matter quasi-particles}

Microcavity polaritons are light-matter quasiparticles that arise from the strong coupling between photons confined in a semiconductor planar microcavity and the excitonic excitations of a material embedded in the microcavity~\cite{Kavokin2007}. A typical structure consists of a cavity spacer of a few tens or hundreds of nanometers in thickness surrounded by an upper and a lower Bragg mirror made out of $\lambda/4$ thick layers of two alternating dielectric materials with different refraction indices (see Fig.~\ref{fig:micropillar}(a)), where $\lambda$ is the operating wavelength of the cavity. The cavity spacer is usually designed for one of the lowest Fabry-Perot modes to be in resonance with the exciton line of a semiconductor material embedded in the central region of the spacer. Photons confined in the cavity excite an exciton (i.e., a bound electron-hole pair), which eventually emits a photon that stays trapped in the microcavity long enough to be re-absorbed, re-emitted and so on. This regime is called "strong light-matter coupling" and is no longer described by exciton and photons independently, but by hybrid quasi-particles called cavity polaritons. The main signature of the light-matter coupling is the splitting of the previously degenerate exciton and photon resonances into two bands called upper and lower polariton branches.

Different materials have been used to implement microcavity polaritons. Strong light-matter coupling has been reported in microcavities based on AlGaAs~\cite{Weisbuch1992}, CdTe ~\cite{Dang1998}, GaN~\cite{AV2003, Christmann2008} and ZnO~\cite{Schmidt2007, Feng2013}, in dielectric microcavities with embedded organic~\cite{Kena-Cohen2010, Plumhof2014} and two-dimensional materials~\cite{Liu2015, Dufferwiel2015b,Lempicka2019} (MoS$_2$, MoSe$_2$, 2D Perovskites etc.), and combinations of them~\cite{Coles2014, Paschos2017, Waldherr2018}, among others.

Polariton quasi-particles can be described as a combination of excitons and photons in the following way: $\left|pol \right\rangle= \alpha \left|ph \right\rangle + \beta \left|exc \right\rangle$, where $\alpha$ and $\beta$ are, respectively, the photon and exciton fractions. Their properties are directly related to those of their constituents and their relative content can be tuned by designing the proper microcavity structure~\cite{Carusotto2013}. Thanks to their photonic part, polaritons eventually escape out of the cavity in the form of photons, which carry all the information on amplitude, phase, energy, polarisation (spin) and coherence of the intracavity photon field. This makes spectroscopy of polaritons in photoluminescence experiments an ideal tool to characterise the polariton bands. Moreover, the vertical confinement in the cavity splits the photon modes into transverse electric (TE) and transverse magnetic (TM) linearly polarised modes. It results in a coupling between the in-plane momentum of polaritons and their polarisation, which can be described as a spin-orbit coupling effect that will be the subject of Sec.~\ref{Sec:spin-orbit}.  In addition, as we will see below, by acting on the photonic part with different techniques~\cite{Schneider2017} it is possible to fabricate one- and two-dimensional lattices.

From the excitonic component, polaritons inherit a "mater" part that has two major effects. First, since excitons are electronic excitations of the quantum wells, they present a real spin, which is sensitive to external magnetic fields via a Zeeman shift. Polariton modes at visible wavelengths are therefore significantly sensitive to magnetic fields, contrary to photons in regular materials. This is crucial to implement topological phases with broken time-reversal symmetry, like Chern insulators~\cite{Nalitov2014b, Karzig2015, Klembt2018}. This will be the subject of Sec.~\ref{Sec:magnetic}. Second, Coulomb interactions between excitons results in significant repulsive interactions between polaritons. In the weak density limit, the interactions result in a $\chi^{(3)}$ Kerr nonlinearity, which can alternatively be described as a polariton-polariton contact interaction. Another crucial property is that polaritons demonstrate bosonic stimulated scattering and gain. These properties of polaritons have allowed the observation of optical parametric oscillation~\cite{Stevenson2000, Savvidis2000}, polariton Bose-Einstein condensation and lasing~\cite{Kasprzak2006}, superfluidity~\cite{Amo2009}, the observaton of Bogoliubov-like excitations~\cite{Kohnle2011,Stepanov2019}, and dark~\cite{Amo2011} and bright solitons~\cite{Sich2011}. They are therefore of potential interest to study nonlinear topological phases, which will be the subject of Sec.~\ref{Sec:nonlinear}.

\subsection*{Polariton lattices}

As mentioned above, one of the main assets of microcavity polaritons to investigate topological phenomena is the possibility of engineering one- and two-dimensional lattices~\cite{Schneider2017}. The polariton platform that has been the most flexible in this aspect is represented by microcavities based on AlGaAs semiconductors. These structures are typically grown using Molecular Beam Epitaxy in two-dimensional wafers and present the highest quality factors of all semiconductor microcavity systems in the strong coupling regime (quality factors above $300~000$ have been reported~\cite{Sun2017b}). By performing electron beam lithography and inductively coupled plasma etching of the microcavity, it is possible to laterally confine photons and, therefore, polaritons, preserving a high enough quality factor at low temperatures (5-10~K).

\begin{figure*}[t!]
\begin{center}
  \includegraphics[width=\textwidth]{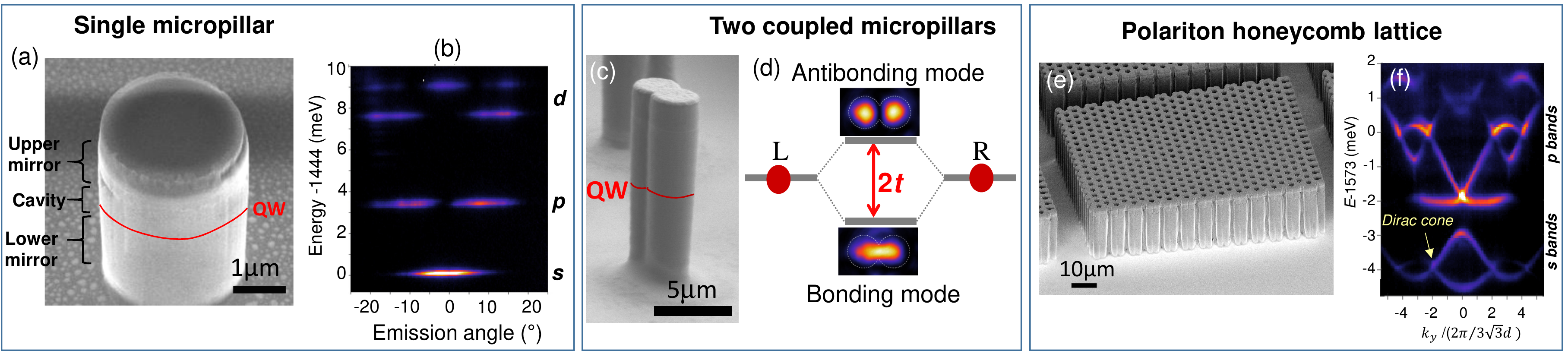}
  \caption{\label{fig:micropillar} (a) Scanning electron microscope image of a AlGaAs-based polariton micropillar. (b) Corresponding spectrum measured in a photoluminescence experiment. (c) Two coupled micropillars and (d) a schematic representation of the bonding and antibonding molecular modes that appear from the coupling of the $s$ modes of each micropillar. (e) Scanning electron microscope image of a polariton honeycomb lattice and (f) the measured spectrum in momentum space.}
   \end{center}

 \end{figure*}

A convenient building block to fabricate lattices is the semiconductor micropillar shown in Fig.~\ref{fig:micropillar}(a). In this microstructure, photons are confined in the vertical direction by the Bragg mirrors and in the horizontal plane by total internal reflection due to the high index of refraction contrast between the semiconductor ($n \sim 3.5$) and air. As polaritons are confined in the three dimensions of space, the spectrum of the micropillar shows a series of discrete $s$, $p$, $d$, ... gapped energy levels, represented in Fig.~\ref{fig:micropillar}(b).

The lithographic mask can be designed to implement two overlapping micropillars, as shown in Fig.~\ref{fig:micropillar}(c). The narrow section between the micropillars acts as a photonic barrier for the coupling of the polariton modes in the two different sites. For instance, if we assume that the lowest energy modes ($s$) of the two individual micropillars have the same energy, the coupled system will display bonding and antibonding modes, separated in energy by $2t$, as sketched in Fig.~\ref{fig:micropillar}(d). Here, $t$ is the coupling strength, and it can be finely tuned by designing the appropriate center-to-center distance between micropillars~\cite{Galbiati2012,Amo2016}.

The technique can be extended to fabricate one- and two-dimensional lattices with high flexibility. One of the advantages of using micropillars as a building block is that in the weak hopping limit, the lattices can be described using a tight-binding approach, in which the on-site energy is controlled by modifying the diameter of the pillars, and the nearest-neighbour hopping by tuning the center-to-center distance between pillars~\cite{Mangussi2020}. An example of a two-dimensional honeycomb lattice is shown in Fig.~\ref{fig:micropillar}(e). The polariton bands can be measured in photoluminescence experiments, in which a non-resonant laser focused in the middle of the lattice excites electrons and holes in the quantum well that relax down forming polaritons that populate the polariton bands. The real-space and angle-resolved detection of the emitted photons using a spectrometer and a CCD camera give access to the real- and momentum-space distributions of polaritons in the lattice. Figure~\ref{fig:micropillar}(f) shows the lowest polariton bands of a lattice similar to that displayed in Fig.~\ref{fig:micropillar}(e). The $s$ bands exhibit Dirac crossings similar to those of electrons in graphene, while the upper $p$-bands present a more elaborate structure~\cite{Jacqmin2014}.

A lower bound for the magnitude of the hoppings that need to be engineered in a typical lattice is given by the polariton linewidth: to experimentally resolve the bands, the bandwidth --proportional to the hopping amplitude $t$-- needs to be larger than the linewidth. This constraint actually places most polariton experiments out of the strict tight-binding limit of weak hopping. This can be readily seen in the asymmetry between the upper and lower bands of both the $s$ and $p$ modes shown in Fig.~\ref{fig:micropillar}(f). Accounting for this asymmetry in a tight-binding model requires the addition of interband couplings, that is, the coupling of $s$ and $p$ modes (and of $p$ and $d$ modes, etc.) in adjacent micropillars, which can be effectively modeled as a next-nearest neighbour correction~\cite{Mangussi2020}. This deviation from a single mode per site model has not prevented the study of a large number of topological effects in polariton lattices, well described by tight-binding Hamiltonians. "Strong" lattice potentials (with bands and bandgaps much larger than the linewidth) were obtained recently in perovskite cavities \cite{Su2020}. Other methods used to create periodic potentials for polaritons include metal film deposition \cite{Kim2013}, acoustic lattices \cite{Cerda2010}, and optical potentials \cite{Berloff2017,Alyatkin2020}.

\section{\label{Sec:Sec2}Topology in scalar polariton lattices}

Implementation of the above mentioned fabrication concepts has led to the experimental investigation of a number of one- and two-dimensional lattice Hamiltonians. Despite the presence of photon losses, which result in a finite linewidth, and the possibility of inducing a local gain (lasing), polariton lattices can be used to study topological phenomena associated to conservative Hamiltonians. As long as the gaps between bands are significantly larger than the linewidth, the concept of topological invariants can still be applied to the bands, and the bulk-edge correspondence connects these invariants to the existence of edge states.

\begin{figure*}[t!]
\begin{center}
  \includegraphics[width=\textwidth]{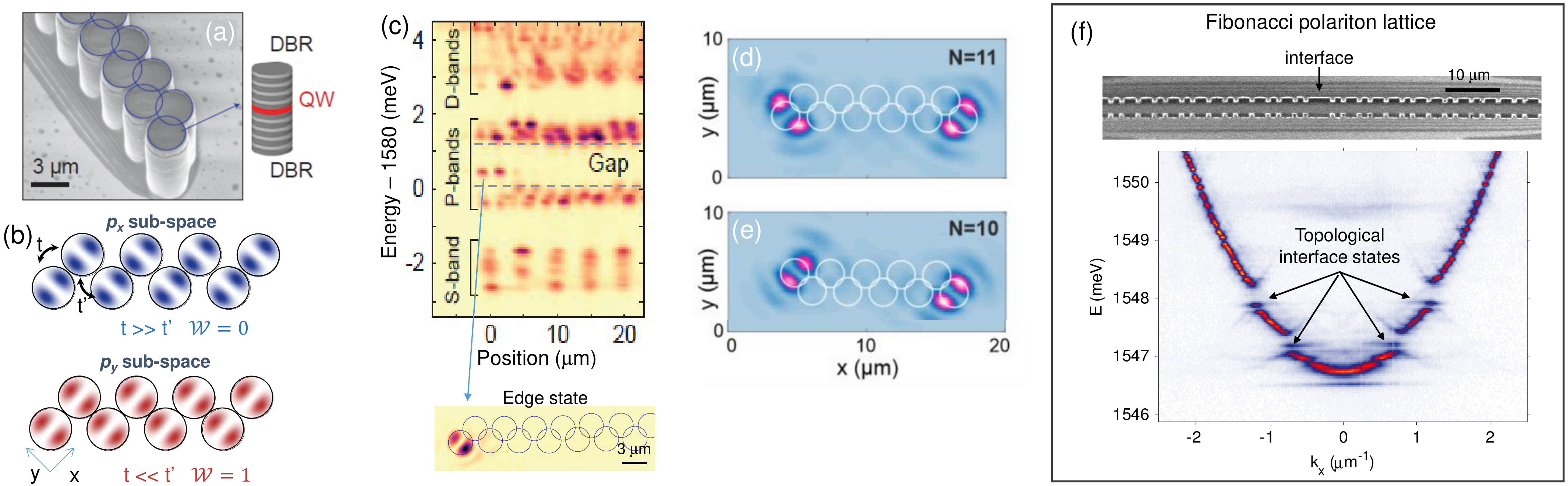}
  \caption{\label{fig:SSH} (a) Scanning electron microscope image of a zigzag polariton lattice implementing the SSH Hamiltonian for $p$ modes. (b) Schematic representation of the $p_x$ and $p_y$ sub-spaces of the lattice showing alternating strong/weak hoppings directly related to the orbital overlap in adjacent micropillars. For the termination represented in the drawing, the $p_x$ sub-space is topologically trivial and the $p_y$ sub-space has a non-zero winding number and shows zero energy states localised at the ends of the lattice. (c) Spatially-resolved spectrum of the lattice displayed in (a). The $p$ bands present a gap with a state localised at the edge. The lower panels shows the real space emission at the energy of this state. It belongs to the $p_y$ sub-space, which is expected to have a topological edge mode for this lattice termination~\cite{St-Jean2017}. (d)-(e) Lasing from $p$ band edge states for different lattice terminations~\cite{Harder2020}. (f) Scanning electron image of a polariton wire with a Fibonacci lateral potential, and corresponding fractal spectrum measured in momentum space showing localised modes in the main gaps~\cite{Baboux2017}.}
   \end{center}
    
 \end{figure*}

One dimensional lattices with topological end states have been implemented in the form of the paradigmatic Su-Schrieffer-Heeger (SSH) Hamiltonian. This lattice model is characterised by a two-sites unit cell with different intra- and inter-cell hopping. It was first proposed for an implementation with polaritons in a zigzag chain \cite{Solnyshkov2016} (following a proposal for a plasmonic structure \cite{Poddubny2014}), and it has been implemented by various groups employing different geometries~\cite{St-Jean2017, Whittaker2019, Harder2020, St-Jean2020}. The SSH Hamiltonian has two bands of eigenvalues separated by a gap. The Hamiltonian is characterised by a nonzero winding number when the inter-cell hopping is larger than the intra-cell one, and a topological edge state in the middle of the gap appears. Using the $s$ modes, the SSH Hamiltonian can be engineered in a linear chain of coupled micropillars with alternating short and long center-to-center distances~\cite{St-Jean2020}.

The SSH Hamiltonian can also be implemented using the first excited modes of each micropillar: the $p$ modes~\cite{St-Jean2017, Whittaker2019, Harder2020}. These modes are doubly degenerate and have a $p_x$ and $p_y$ geometry. To realise the SSH Hamiltonian with $p$ modes, a zigzag chain of coupled micropillars with identical center-to-center distance needs to be fabricated (Fig.~\ref{fig:SSH}(a)). The zigzag geometry alternates strong and weak couplings for each of the $p_x$ and $p_y$ sub-spaces of the lattice, realising two independent copies of the SSH Hamiltonian (Fig.~\ref{fig:SSH}(b)). The precise termination of the chain sets the dimerization (either $t>t'$ or $t<t'$) of each sub-space and, determines which sub-space will show an edge states. Figure~\ref{fig:SSH}(c) displays the real space spectrum of the lattice shown in (a). The $p$ bands present a gap and a state localised at the end of the chain, belonging to the topologically nontrivial $p_y$ sub-space. In the other possible termination configuration, the edge mode appears in the $p_x$ sub-space, as demonstrated in Fig.~\ref{fig:SSH}(d)-(e)~\cite{Harder2020}. Lasing from these edge modes provided the first signature of lasing in a topologically protected mode~\cite{Solnyshkov2016,St-Jean2017}. This mode is well isolated in the middle of the gap, and its energy is insensitive to weak disorder in the hoppings.

The flexibility in the design of the local polariton landscape has permitted the investigation of the topological properties of one-dimensional aperiodic potentials. Figure~\ref{fig:SSH}(f) shows a scanning electron microscope image of a one-dimensional polariton lattice whose lateral confining potential follows a Fibonacci sequence~\cite{Tanese2014}. Its associated spectrum presents a fractal structure with series of gaps characterised by a topological invariant. The presence of edge states associated to this invariant and the value of the invariant have been directly measured in structures with an interface between two mirror images of a Fibonacci potential~\cite{Baboux2017}.

\begin{figure*}[t!]
\begin{center}
  \includegraphics[width=\textwidth]{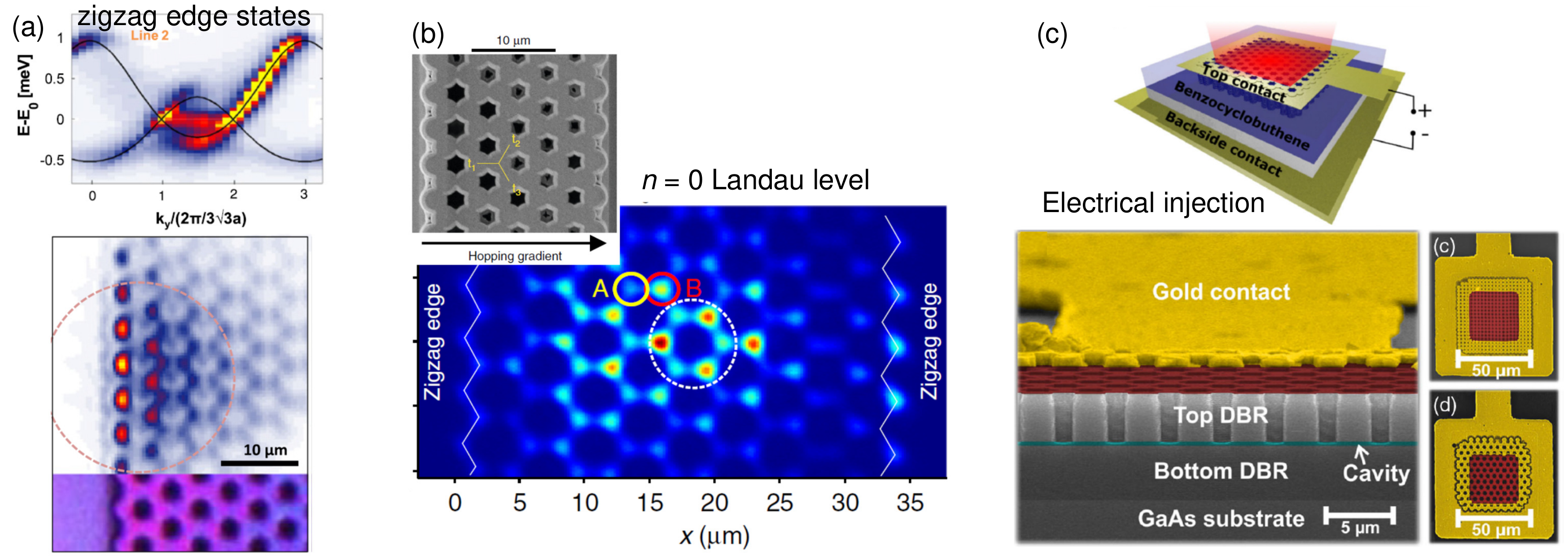}
  \caption{\label{fig:honeycomb} (a) Top: Momentum resolved emission of a honeycomb lattice excited at the zigzag edge. The flat line that joins together the two Dirac points corresponds to the edge states. Bottom: real space emission at the energy of the of the zizag edge states. From Ref.~\cite{Milicevic2015}. (b) Scanning electron microscope image of a honeycomb lattice subject to artificial uniaxial strain along the horizontal direction, and real space emission at the energy of the $n=0$ Landau level in a similar lattice. The A-B sublattice asymmetry is a distinctive feature of the lowest order Landau level in honeycomb lattices subject to an artificial magnetic field. From Ref.~\cite{Jamadi2020}. (c) Scheme and scanning microscope electron images of a polariton honeycomb lattice with electrical injection, in which polariton lasing was observed. From Ref.~\cite{Suchomel2018}.}
   \end{center}
    
 \end{figure*}

In two dimensional lattices, the first observation of edge states connected to the topology of a polariton lattice corresponds to the $s$ bands in honeycomb lattices~\cite{Milicevic2015} (see Fig.~\ref{fig:honeycomb}(a)). In the absence of any external magnetic field, these lattices present time-reversal, particle-hole and sublattice symmetry. They fall into the BDI class (chiral orthogonal) in the classification of topological phases~\cite{Schnyder2008}, and they cannot be described by non-trivial $\mathcal{Z}$ or $\mathcal{Z}_2$ invariants (Chern numbers). In fact, the two $s$-bands in this lattice, which mimic the $\pi$ and $\pi^*$ bands of graphene, are gapless in the bulk: they touch at the Dirac points, as shown in Fig.~\ref{fig:micropillar}(f). Nevertheless, the lattice shows edge states at the Dirac energy, and they are of topological origin in the sense that they can be related to the existence of bulk invariants: a winding number defined in a sub-space of the lattice Hamiltonian~\cite{Ryu2002, Delplace2011}. Contrary to Chern and topological insulators, in which the existence of edge modes is independent of the geometry of the edge, the existence of edge states in the Honeycomb lattice is dependent on the precise geometry of the edges, which sets the value of 0 or 1 of the winding number. The existence of edge states for certain edge geometries (zigzag and bearded) was noticed in early works on graphene~\cite{Fujita1996, Nakada1996, Kohmoto2007} and it was later extended to any kind of edge~\cite{Delplace2011}. Using polariton honeycomb lattices, it has been possible to prove the existence of these edge states experimentally and to characterise their momentum distribution for zigzag and bearded edges~\cite{Milicevic2015}. Interestingly, edge states of similar nature have been observed in the $p$ bands of polariton honeycomb lattices~\cite{Milicevic2017}. More recently, it has been possible to directly measure the winding numbers associated to the bulk bands in photoluminescence experiments~\cite{St-Jean2020}. By measuring the real space intensity distribution for selected in-plane momenta it is possible to compute the so-called "mean chiral displacement" which provides a measurement of the winding number in lattices with chiral symmetry~\cite{Mondragon-Shem2014, Cardano2017, Maffei2018}.

The possibility to control the site-to-site hopping amplitude in lattices of coupled micropillars has permitted the exploration of lattice models that imitate the effect of spatial strain. In honeycomb lattices, the presence of uniaxial strain has been shown to result in the merging of the Dirac cones and the opening of a gap. In polariton lattices, this situation can be realised by increasing the $s$ mode hopping amplitude of one of the three nearest-neighbour connections of each micropillar. The merging of the Dirac cones results in a semi-Dirac dispersion, with highly anistoropic transport properties~\cite{Real2020}, and when the gap opens, edge states in bearded terminations disappear~\cite{St-Jean2020}, as reported in previous experiments in coupled waveguides~\cite{Rechtsman2013c} and microwave resonators~\cite{Bellec2014}.

When the strain in the honeycomb lattice is designed with a spatial gradient (i.e., the hopping amplitude varies in space along a particular direction -- upper panel in Fig.~\ref{fig:honeycomb}(b)), a synthetic magnetic field appears close to the Dirac cones~\cite{Guinea2010, Salerno2015}. The field is directly proportional to the spatial gradient of hopping amplitudes, and it results in an artificial magnetic field of opposite sign at the K and K' Dirac cones. This geometry was first realised in photonics in coupled waveguides~\cite{Rechtsman2013}. A recent realisation in polariton lattices~\cite{Jamadi2020} has allowed accessing experimentally the real space distribution of the wavefunction of the Landau levels associated to the artificial field (Fig.~\ref{fig:honeycomb}(b); see also Ref.~\cite{Bellec2020}), and the observation of helical edge states~\cite{Salerno2017}.

Cavity polaritons allow also accessing novel topological regimes that cannot be described by invariants defined in conservative systems. Recently, polariton lattices based on the interplay of gain and interactions with the excitonic reservoir injected by a nonresonant laser have been reported in a geometry based on the SSH lattice~\cite{Pickup2020synthetic}. A topological gap and interface states appear in this lattice whose origin is purely non-Hermitian. In a different work, it was shown theoretically that topological phase transitions can be driven by the pattern of a nonresonant pump laser in a lattice of micropilars with constant hopping between adjacent sites~\cite{Comaron2020nh}. These works, very much related to exceptional points, open interesting perspectives in the study of genuinely non-Hermitian topology in polariton lattices.

Going beyond microcavities with standard quantum wells, a recent theoretical work has demonstrated a peculiar interplay of the strong coupling with the topology of a lattice of localised dipoles~\cite{Downing2019}, in which the usual bulk-edge correspondence present in electronic systems breaks down.


\begin{figure*}[t!]
\begin{center}
  \includegraphics[width=\textwidth]{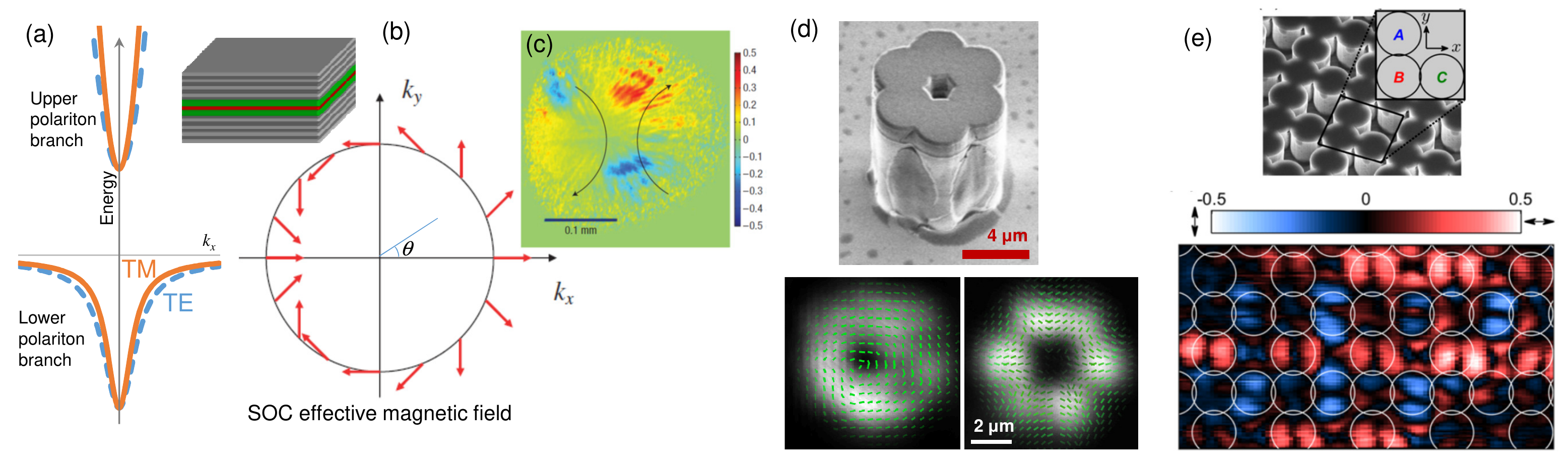}
  \caption{\label{fig:SOC} (a) Schematic dispersion of the polariton branches in a planar dielectric microcavity, showing the TE-TM splitting. (b) Direction of the effective magnetic field acting on the polarisation pseudospin as a function of $k_x$ and $k_y$ for a given value of the norm of the in-plane momentum $k$. (c) Spatial polarisation pattern of polaritons propagating radially, resonantly injected with linear polarisation~\cite{Leyder2007}. The color scale shows the degree of circular polarisation. The formation of the pattern is a consequence of the polarisation pseudospin precession due to the pressence of the effective magnetic field shown in (b). (d) Scanning electron microscope image of a benzene photonic molecule, and the measured spatial polarisation texture of two molecular s modes~\cite{Sala2013}. The green lines show the measured plane of linear polarisation at each point in space.  (e) Scanning electron image of a Lieb lattice etched in a semiconductor microcavity and the polarisation pattern of flat band modes originating in the SOC~\cite{Whittaker2018}.}
   \end{center}

 \end{figure*}

\section{\label{Sec:spin-orbit}Spin-orbit coupling in polariton lattices}

So far, we have neglected the polarisation of polaritons. The photonic component of polaritons has a polarisation degree of freedom with two spin projections, and the excitons to which these photons are coupled, also have two spin projections $J_z=\pm 1$. Excitonic modes with spin projections $J_z=0,\pm2$ are also possible in the quantum wells but are not coupled to light; they correspond to dark exciton states and we will not consider them here. In most of the above mentioned realisations, the polarisation degree of freedom does not play an important role and the observed phenomena are independent of the polarisation polaritons. As we will see now, the polarisation properties of microcavities present quite interesting phenomenology.

Photon eigenmodes in two-dimensional optical systems present transverse electric (TE) and transverse magnetic (TM) polarisations~\cite{Bliokh2015}. This splitting is significant everywhere in the two-dimensional momentum space except near $k=0$ (normal incidence) where both the magnetic and electric field of transverse waves stand in the plane of the two-dimensional cavity. 
Figure~\ref{fig:SOC}(a) shows a scheme of the polariton dispersion of a planar microcavity made out of semiconductor Bragg mirrors. Two polariton branches are visible, each of them resolved in linear polarisation. At in-plane momentum $k=0$, both polariton branches show no polarisation splitting if linear birefringence is absent. However, for higher momenta, a polarisation splitting between linear polarisations parallel and perpendicular to the direction of the in-plane momentum $k_x$ is apparent in both branches. These polarisation directions correspond to the TE and TM eigenmodes, and the splitting is known as the TE-TM splitting. This TE-TM 
splitting appears in any inhomogeneous system, in presence of a spatial gradient in any optical parameter, which allows to define the two transverse polarizations~\cite{Landau8, Bliokh2015}.
In microcavities, the main source of this splitting is the polarisation dependence of the reflectivity of light in the dielectric layers forming the Bragg mirrors, which depends on the angle of incidence~\cite{Panzarini1999b} and, consequently, on the value of the in-plane momentum $k$. The splitting is strongly reduced for the lower polariton branch at large values of the momentum, for which polaritons are mostly excitonic, because the exciton longitudinal-transverse splitting is usually much smaller than the photonic TE-TM splitting. The situation is very different in cavities with two-dimensional transitional metal dichalcogenide active layers, where the TE-TM splitting 
grows continuously with the wave vector even beyond the light cone \cite{Glazov2014,Vasilevskiy2015,Bleu2017tmd}.

The linear polarisation TE-TM splitting can be modelled as an effective magnetic field $\mathbf{\Omega (k)}$ acting on the $1/2$ pseudospin that describes the two circular polarisation components of light with a given in-plane momentum $\mathbf{k}$. This pseudospin can be represented by a three-dimensional vector in the Poincar\'e sphere, in which the poles indicate purely circular polarisations and the x-y plane describes linear polarisation along all possible directions within the plane of the cavity. The effective magnetic field acting on the polarisation pseudospin takes the form~\cite{Shelykh2004c, Kavokin2005}:

\begin{equation}
\mathbf{\Omega (k)} = \left( \frac{\Delta_{TE-TM}(k)}{\hbar} \cos{2\theta}, \frac{\Delta_{TE-TM}(k)}{\hbar} \sin{2\theta}, 0 \right)
\label{eq:TETMsplt}
\end{equation}

\noindent where $\theta$ is the in-plane angle of propagation of polaritons in the cavity sketched in Fig.~\ref{fig:SOC}(b) ($\theta=\arctan k_y / k_x)$, and $\Delta_{TE-TM}(k)$ is the value of the linear polarisation splitting for a given value of $k$. Note that $\Delta_{TE-TM}$ does not only depend on $k$ but also on the design of the microcavity structure. Indeed, by modifying the thickness of the dielectric layers in the Bragg mirrors with respect to that of the cavity spacer, $\Delta_{TE-TM}$ can be positive, negative or zero~\cite{Panzarini1999b}.

The dependence of the TE-TM splitting on the propagation direction in planar microcavities can be seen as a form of spin-orbit coupling (SOC). Its main consequence for the balistic coherent propagation of polaritons is the precession of the polariton pseudospin in a phenomenon know as optical spin Hall effect~\cite{Kavokin2005,Langbein2007, Leyder2007,Langbein2007,Amo2009a, Kammann2012,Lundt2019tmd} (see Fig.~\ref{fig:SOC}(c)). In combination with additional polarisation splittings linked with linear birefringence, it gives rise to intricate non-Abelian gauge fields in planar structures~\cite{Tercas2014,Fieramosca2019}.

When considering a single micropillar, the TE-TM SOC results in a fine structure and elaborate spin textures for modes with orbital momentum different from zero~\cite{Dufferwiel2015}. In photonic molecules and lattices, the TE-TM SOC is present also for molecular orbitals and bands formed from the coupling of the fundamental ($s$) modes of each individual pillar. It can be described as a polarisation dependent hopping in tight-binding models, because it induces different confinement barriers and different effective masses for polaritons with linear polarisations oriented parallel or perpendicular to the line connecting the two adjacent pillars, which means different tunneling coefficients~\cite{Sala2013, Nalitov2014b,Nalitov2015, Mangussi2020}. The coupling of orbital and polarisation degrees of freedom has been used to engineer benzene-like photonic molecules where lasing can occur in modes presenting interesting polarization textures~\cite{Sala2013}. Remarkably, these strucutre can also lase in modes with a net orbital angular momentum whose chirality can be tune via the polarization of an off-resonant pump~\cite{CarlonZambon2019}. 
In polariton lattices, the interplay of the lattice geometry and the SOC modifies the effective form of the SOC~\cite{Klembt2017, Whittaker2018} (see Fig.~\ref{fig:SOC}(e)). For instance, in a honeycomb lattice close to the Dirac cones, the SOC takes the form of a Dresselhaus field~\cite{Nalitov2015, Whittaker2020}.

Even in one-dimensional polariton chains, the SOC plays an important role, leading to a dimerization of a zigzag chain of polariton micropillars and rendering it equivalent to a SSH chain \cite{Solnyshkov2016, Whittaker2019}. This provides an extra topological protection to the polarization domains formed during polariton condensation via the Kibble-Zurek mechanism, which should allow observing the corresponding scaling in continuous wave experiments.

A completely different approach to engineer spin-orbit coupling for polaritons has been realised realised using a monoatomic layer of a transition metal dichalcogenides (WS$_2$ and MoSe$_2$) deposited on top of a photonic crystal. 
The crystal realizes the topological insulating phase for photons proposed by Wu and Hu in Ref.~\cite{Wu2015b}, in which topological interface modes with a well defined circular polarisation are present. The exciton transitions in transition metal dichalcogenide strongly coupled to the photonic modes of the crystal resulting in polaritons with the topological properties of the underlying photonic crystal~\cite{Liu2020,Li2020tmd}. 

\section{\label{Sec:magnetic}Topology of microcavities under magnetic field}

\subsection*{Zeeman splitting and spin orbit coupling in microcavity polaritons}



One of the most advantageous features of polaritons with respect to other optical systems is the giant Faraday effect they show in the presence of an external magnetic field~\cite{Kavokin1997}. The origin of the Faraday effect is the partial matter component of polaritons: the polariton modes present a strong Zeeman splitting at optical frequencies inherited from the exciton Zeeman splitting~\cite{Snelling1992}. This splitting has been found to be larger than the polariton linewidth under external fields of a few Tesla in GaAs- and CdTe-based microcavities~\cite{Walker2011, Larionov2010, Sturm2015,Pietka2015, Mirek2017}. As we will describe below, the polariton Zeeman splitting in combination with the TE-TM SOC allow the implementation of a polariton Chern insulator.

The first study of the topological Berry phase in polariton structures with TE-TM SOC and magnetic field dates back to 2009~\cite{Shelykh2009}. In this work, the Berry phase accumulated along a circular path in reciprocal space with non-zero Berry curvature was suggested to be used as the control parameter of an interferometer of the Aharonov-Bohm type. This work introduced the simplest Hamiltonian that captures the essence of 
polariton physics in the presence of a magnetic field:
\begin{equation}
\label{eq:magpolaritons}
\hat H = \left( {\begin{array}{*{20}{c}}
{\frac{{{\hbar ^2}{k^2}}}{{2{m^*}}} + \frac{{{\Delta _Z}}}{2}}&{\frac{{{\Delta _{TE-TM}}\left( k \right)}}{2}{e^{ - 2i\varphi }}}\\
{\frac{{{\Delta _{TE-TM}}\left( k \right)}}{2}{e^{2i\varphi }}}&{\frac{{{\hbar ^2}{k^2}}}{{2{m^*}}} - \frac{{{\Delta _Z}}}{2}}
\end{array}} \right)
\end{equation}

where we have chosen the pseudospin basis of the two circular polarisation projections $\sigma +$ and $\sigma -$ of the polariton field, and $\Delta_Z$ is the polariton Zeeman splitting. This Hamiltonian can be represented as an effective field $\mathbf{\Omega}=(\Delta_{TE-TM}\cos 2\varphi,\Delta_{TE-TM}\sin 2\varphi,\Delta_Z)$ acting on the polariton pseudospin in the Poincar\'e sphere. The field is a superposition of the TE-TM effective field \ref{eq:TETMsplt} and the Zeeman field.

Using Eq.~\ref{eq:magpolaritons} it has been shown that the polariton branches in a planar microcavity in the presence of an external magnetic field and TE-TM SOC contain non-local Berry curvature~\cite{Bleu2018a, Bleu2018b}, later evidenced in experiments~\cite{Gianfrate2020}. Because the TE-TM splitting grows quadratically for small wave vectors ($\Delta_{TE-TM}\sim k^2$), the Berry curvature in a planar microcavity exhibits a ring-like distribution in reciprocal space, with a maximum approximately corresponding to the region where the TE-TM splitting becomes comparable with the Zeeman splitting $\Delta_{TE-TM}(k)\approx \Delta_Z$. Similar Berry curvature distributions have also been found in microcavities with linear birefringence~\cite{Kokhanchik2020} and optical activity~\cite{Ren2019}, and the control of the distribution of the Berry curvature via the exciton-photon detuning in a multiband system (a perovskite cavity with a thick active region) has been demonstrated in \cite{Polimeno2020}.


Polariton bands with locally non-zero Berry curvature can also be obtained using optically active materials or {\textit{via}} emergent optical activity shown to appear in cavities with strongly birefringent materials \cite{Rechcinska2019,Kokhanchik2020}. Naturally, the maxima of the Berry curvature are associated in this case with the regions where the effective field changes rapidly, that is, with anticrossings in the polariton or photonic branches \cite{Ren2019}.

\begin{figure*}[t!]
\begin{center}
  \includegraphics[width=0.7\textwidth]{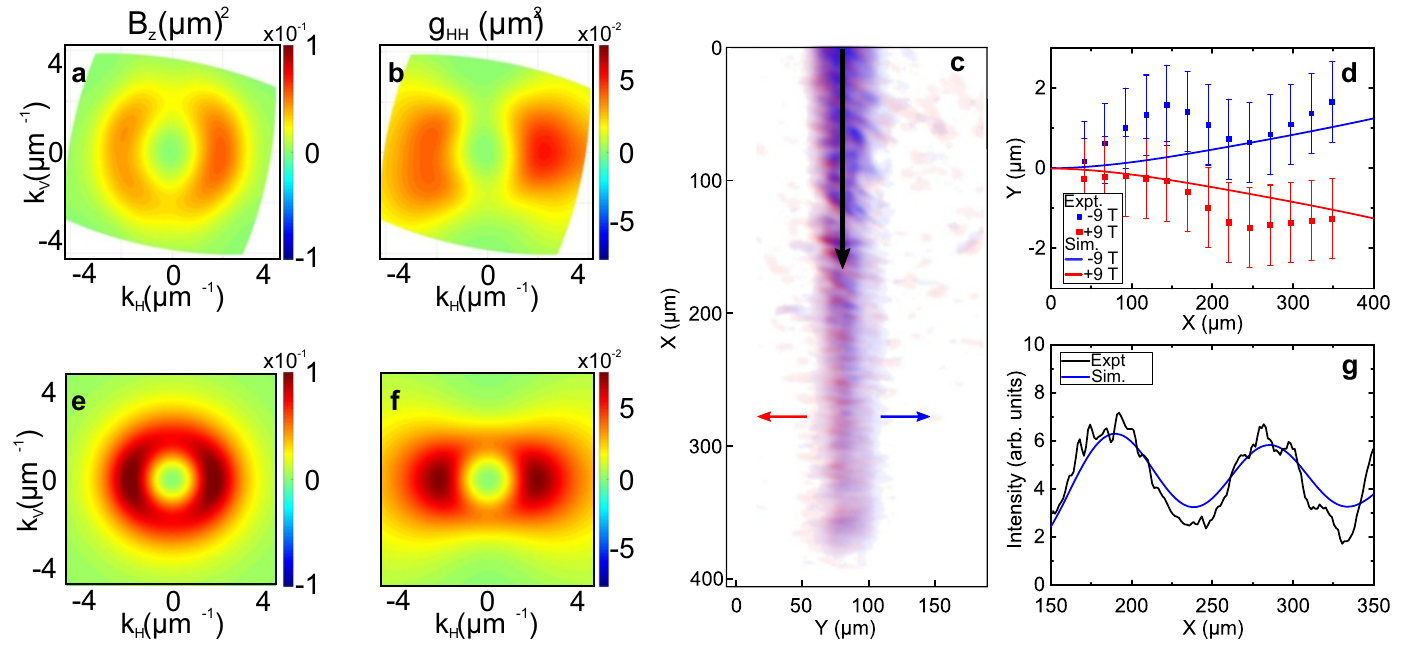}
  \caption{\label{fig:qgt} (a,e) Berry curvature (experiment, theory), (b,f) quantum metric tensor element $g_{HH}$ (experiment, theory) for the polariton modes in a planar microcavity. (c) Experimental spatial images of the anomalous Hall drift for opposite magnetic fields; (d) Trajectory deviation due to the anomalous Hall effect (experiment and theory based on the Berry curvature from panel (a)); (g) intensity oscillations, whose contrast is determined by the quantum metric (from panel (b)). From Ref.~\cite{Gianfrate2020}}
   \end{center}

 \end{figure*}

Finally, we point out a recent implementation of an artificial gauge field for polaritons realized by applying a combination of a magnetic field and an electric field to a semiconductor microcavity~\cite{Lim2017}. Using the electric polarizability of excitons, it is possible to induce a dipole for $k \neq 0$ polaritons propagating under a magnetic field. Additionally, by applying an electric field in a direction orthogonal to the magnetic field direction, the polariton dispersion gets modified: new terms appear in the associated Hamiltonian that are directly proportional to the wavevector $k$, and that can thus be recast as a vector potential. This scheme, implemented on a planar cavity, has enabled observing geometric phase accumulations of up to 0.25~{\rm rad} after about 9~{\rm $\mu$m} polariton propagation~\cite{Lim2017}. Such electrically tunable artificial gauge potential could readily be implemented in polariton lattices, where it could constitute another useful way to achieve time reversal symmetry breaking.

\subsection*{Anomalous Hall drift and the quantum geometric tensor}

While the existence of chiral edge states with topological protection requires the system to represent a closed manifold with a non-trivial Berry curvature and to exhibit an energy gap, other effects, determined by locally-nontrivial Berry curvature can be observed and even used for practical applications in a larger class of systems without these stringent conditions.

The most prominent example is the anomalous Hall effect predicted in 50s \cite{Karplus1954} and interpreted as a consequence of the Berry curvature in the 90s \cite{Sundaram1999}. The anomalous Hall velocity appears in addition to the ordinary group velocity when a wave packet traverses a region of non-zero Berry curvature in the reciprocal space. This effect is the cornerstone of valleytronics (where it is called valley Hall effect \cite{Mak2014}), since valleys often exhibit a localised non-zero Berry curvature even if the system is topologically trivial as a whole. In microcavity systems, this effect has been studied theoretically in a honeycomb lattice in which a photonic wavepacket is subject to an artificial acceleration~\cite{Ozawa2014, Cominotti2013}

Interestingly, the Berry curvature is only part of a more general object called the quantum geometric tensor \cite{Provost1980}. Indeed, the quantum geometric tensor has two components: the Berry curvature and the quantum metric, whose role in physical phenomena is only starting to be understood. 
In particular, it was shown that for any finite-duration experiment, first and second order corrections to the anomalous Hall effect are determined by the quantum metric \cite{Gao2014,Bleu2018a}. In general, the quantum metric determines the overlap between the different quantum states in an adiabatic evolution, and for this reason it is currently used in quantum information theory. It is expected to be particularly important in non-Hermitian systems, including polariton modes in microcavities \cite{Richter2019,Solnyshkov2020nh1}.

The link between the Berry curvature distribution in reciprocal space (band geometry) and the anomalous Hall drift in polariton systems has been experimentally demonstrated in a quantitative way in a planar microcavity~\cite{Gianfrate2020}. Both the Berry curvature and the quantum metric have been measured experimentally, as shown in Fig.~\ref{fig:qgt}(a,b) (theoretical distributions -- e,f).  In the same work, the polariton anomalous Hall effect (panels c,d) was experimentally observed, the measured trajectory being in agreement with the theoretical calculations using the independently measured Berry curvature as input parameter. Figure~\ref{fig:qgt}(g) shows intensity oscillations as a function of the polariton propagation in the polarisation opposite to that of the injected polaritons. Their contrast is determined by the non-adiabatic processes, which are accounted for by the non-zero value of the quantum metric.

\subsection*{Polariton Chern insulators}



Photon Chern insulators were first proposed by Haldane and Raghu \cite{Haldane2008} and observed soon after by the group of Soljacic \cite{Wang2009}. One key element of this proposal was the breaking of time reversal symmetry for photons. In practice this required the use of gyromagnetic materials, which has limited for a long time the observation of this effect to the microwave domain. In the original implementation of Wang and coworkers~\cite{Wang2009}, bands with a nontrivial Chern number appeared in a two-dimensional lattice of rods sandwiched between two metallic plates. 

A crucial aspect in the formation of the Chern bands, which was only highlighted later, is the role of the transverse nature of the confined electromagnetic waves~\cite{Bliokh2015,Nalitov2014b,Bardyn2014}. Indeed, as discussed in the previous section, the transverse nature of eigenmodes is responsible for an intrinsic polarisation splitting, which can be seen as an intrinsic chirality \cite{Bliokh2015}.
The recipe to implement Chern phases in microcavities is based on that concept~\cite{Nalitov2014b,Bardyn2014}: one needs to combine the TE-TM polarisation splitting with the polariton Zeeman splitting, which in the case of microcavity polaritons is significant at visible wavelengths. Actually, the specific form of the polarisation splitting in microcavities as a function of in-plane momentum determines the value of the Chern number of the topological bands at low magnetic fields.  

\begin{figure*}[t!]
\begin{center}
  \includegraphics[width=\textwidth]{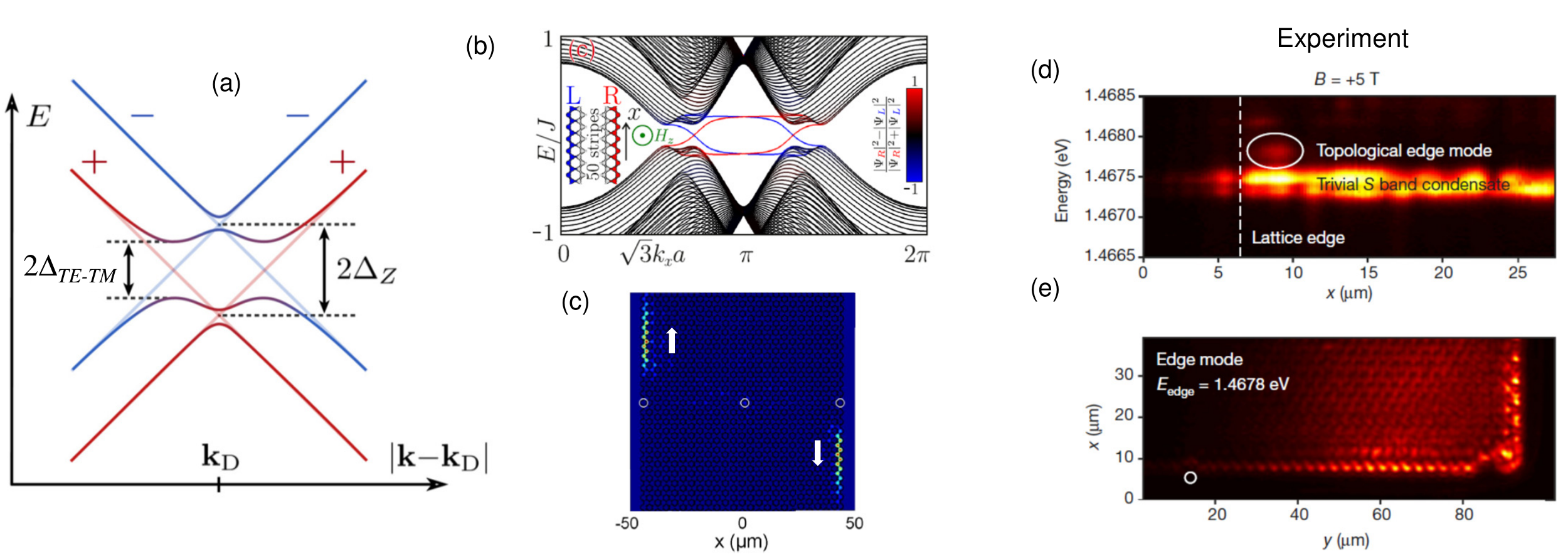}
  \caption{\label{fig:Chern} (a) Scheme of the gap opening at a Dirac crossing due to the combination of external magnetic field resulting in a Zeeman splitting $\Delta_Z$ betwwen bands with $\sigma+$ and $\sigma-$ polarisation, and the TE-TM splitting $\Delta_{TE-TM}$. From Ref.~\cite{Bardyn2014}. (b) Tight-binding simulation of the bands in a honeycomb lattice ribbon with zigzag edges showing the dispersion of edge states associated to the non-trivial gap~\cite{Nalitov2014b}. (c) Simulated evolution of two wavepackets created at the left and right edge-mode bands in a honeycomb ribbon. The snapshot shows the position of the wavepackets 100~ps after injection at the points marked by circles. The chirality of the edge states is demonstrated by the opposite directions of propagation at opposite edges~\cite{Nalitov2014b}. (d) Measured spectrum of a polariton honeycomb lattice in a magnetic field of 5~T showing polariton lasing in bulk modes and emission from edge modes. (e) Measured real-space emission at the energy of the topological edge mode showing high intensity coming from the edges of the lattice~\cite{Klembt2018}.}
   \end{center}

 \end{figure*}

The first theoretical proposals for polaritons employed the tight-binding approximation in which the TE-TM SOC is described as a polarisation-dependent tunneling~\cite{Nalitov2014b, Galbiati2012, Sala2013}.
In combination with the Zeeman splitting, the polarization-dependent tunneling gives rise to a topological gap at band crossings of a lattice, for instance, at the Dirac points of a honeycomb lattice (Fig.~\ref{fig:Chern}(a)). The topological gap is associated with the presence of chiral edge states whose group velocity is linked with the edge (one-way states, with opposite velocity at opposite edges) (Fig.~\ref{fig:Chern}(b),(c)).

A phenomenological way of understanding the opening of the gap at the Dirac crossings is represented in Fig.~\ref{fig:Chern}(a). The Zeeman effect acting on polaritons splits the original two bands into four bands with psudospin (circular polarisation) $\sigma+$ and $\sigma-$. The four bands cross in pairs at four different points. The TE-TM SOC mixes the $\sigma+$ and $\sigma-$ bands at the crossings, resulting in four anti-crossings that open a central gap. In the limit of TE-TM splitting $\Delta_{TE-TM}$ being much smaller than the Zeeman splitting $\Delta_{Z}$, the magnitude of the topological gap is dominated by $\Delta_{TE-TM}$.

Let us stress again that the behavior of the TE-TM SOC in microcavities and in photonic crystals is very different. In microcavities, the SOC is zero at $k_{\parallel}=0$. It can therefore be considered as a small correction to the dispersion at low wave vectors. On the contrary, in photonic crystals, the branches of different polarisations are usually strongly split: in this case, the SOC is much larger than the other terms in the Hamiltonian (even the terms induced by the potential of the lattice). In polariton graphene, with realistic parameters, the band Chern number is $\pm 2$, and the phase realised is not exactly the one initially described by Haldane and Raghu \cite{Haldane2008}. When increasing either the TE-TM SOC or the Zeeman splitting, a phase transition occurs associated to the closure of the gap at the M-point~\cite{Bleu2016,Bleu2017x}. At the transition, Chern numbers change sign and values, passing from $\pm 2$ to $\mp 1$. This second phase, which was also predicted for the topological gap of the $p$-bands \cite{Zhang2019pband} is the analoguous of the phase found by Haldane and Raghu.
When this transition occurs the number of topologically protected chiral states at each edge, determined by the Chern number, also changes, and the edge modes change their propagation direction. Since this transition can occur by controlling the magnitude of the Zeeman splitting, it can be ultimately used for optically controlled microscopic, topological, optical isolators taking advantage of the polarisation-dependent interactions of polaritons \cite{Solnyshkov2018}, discuseed in detail in Sec.~\ref{Sec:nonlinear}. The efficiency of such isolators depends on the exponentially small overlap of the edge states localised at the opposite edges. If these states exhibit a non-negligible overlap, they become coupled. The suppression of this coupling was considered in \cite{Zhang2018inh}. However, this coupling can also potentially lead to applications, as considered in some recent works \cite{Zhang2018r}. Other possible applications of edge states in specially engineered polariton topological insulators include spin filtering \cite{Mandal2019} (as compared with valley filtering proposed in \cite{Bleu2017}).

In parallel, the work of Bardyn and coworkers considered the limit of strong TE-TM SOC~\cite{Bardyn2014}: the coupling of a single photonic TE branch with circular-polarised excitonic states can be seen as exhibiting a topologically-nontrivial winding, precisely because the single TE branch is already topologically-nontrivial. The two polariton branches formed for each spin in this case exhibit a non-zero Berry curvature, and a topological gap can be opened by using a periodic potential (for excitons or for photons \cite{Karzig2015}). Other possibilities were also explored, such as the coupling of photonic Dirac cones with excitons \cite{Yi2016}. For example, topological insulators based on magnetic dots were proposed \cite{Sun2019}. Floquet topological polariton lattices created by time-dependent potentials were also considered \cite{Ge2018f}. Polariton topological insulators based on lattices other than the honeycomb lattice were also studied, for example, the Lieb polariton insulator \cite{Li2018l}. 

Experimentally, evidence of the formation of a polariton Chern insulator has been reported by Klembt and coworkers in Ref.~\cite{Klembt2018}. 
In that work, the authors measured the dispersion in presence of an external magnetic field. A topological edge mode was observed in the emission spectrum (Fig.~\ref{fig:Chern}(d)) and its localization was confirmed by the analysis of the real space emission (Fig.~\ref{fig:Chern}(e)). The ratio of the intensities at the $K$ and $K'$ points was shown to depend on the sign of the magnetic field, which was interpreted as a signature of chirality. Very interestingly, the employed platform supports electrical injection (see Fig.~\ref{fig:honeycomb}(c)) and had been used to demonstrate electrically-driven polariton lasing from states at the $\Gamma$ points~\cite{Suchomel2018}, similar to the case of optical pumping in \cite{Jacqmin2014}.

\section{\label{Sec:nonlinear}Topology and nonlinearities in microcavity polaritons}

\subsection*{Nonlinear effects in topological edge states}

The first crucial question concerning non-linear effects is the possibility of topological lasing, that is, lasing from the topologically protected edge states found in the linear regime. 
Polaritons are particularly suited to the exploration of these effects thanks to the extensive experience of the polariton community on lasing in planar and confined microcavity structures.
Indeed, lasing in topologically protected modes was the focus of some of the earliest works in topological polaritonics, both theoretical~\cite{Solnyshkov2016} and experimental~\cite{St-Jean2017}.

The system studied in these works was the zigzag chain of pillars, presented in Sec.~\ref{Sec:Sec2}, which supports localised edge modes of topological origin (see also Ref.~\cite{Harder2020}). Moreover, in polariton lattices, condensation (lasing) occurs preferentially in localised states that exhibit a higher overlap with the pumping laser than the propagating states, as experimentally shown in several works \cite{Tanese2013,Jacqmin2014}. Indeed, repulsive interactions between polaritons expel them out of the pumping region if they belong to propagative modes \cite{Wertz2010}, whereas the localized states cannot be expelled and thus show a better overlap, and a higher gain. As a result, lasing preferentially occurred on the edge modes of the zigzag chain rather than in the bulk modes. 
Almost simultaneously to the publication of these works, several reports demonstrated one- and two-dimensional topological lasing in non-polaritonic platforms \cite{Parto2018,Zhao2018, Bahari2017,Bandres2018, Harari2018}. 

In two-dimensional polariton Chern insulators under an external magnetic field~\cite{Klembt2018}, topological edge states were observed under strong pumping conditions. A polariton lasing mode was formed in the bulk states of the lattice at the K and K' points, which reduced the linewidth of the bulk modes and facilitated the observation of the edge sates (see Fig.~\ref{fig:Chern}(d)). 

Recent theory works have addressed the conditions for polariton lasing from the edge states in two-dimensional systems and studied the stability of such lasing~\cite{Kartashov2019, Secli2019}, showing that depending on the gain profile topological lasing can be subject to absolute or convective instabilities~\cite{Secli2019}, and present Kardar-Parisi-Zhang scaling properties~\cite{Amelio2019}.

A different approach to the study of nonlinear behaviour of topological polariton lattices is based on the polariton-polariton interactions. Thanks to their matter (excitonic) component, polaritons present significant particle-particle interactions. They can be modeled as contact, repulsive interactions, of the Kerr type in the context of optics and they can be studied in the context of topological lattices. A few works have studied the stability of topological states in the presence of polariton interactions \cite{Kartashov2016b, Ma2020}, and predicted that the topological edge states should be unstable in certain conditions~\cite{Longhi2018b}. Linear and nonlinear interface states in polariton topological insulator junctions were considered in \cite{Zhang2019inter}. The coupling of multiple edge states and the formation of topological Bragg solitons were studied in \cite{Zhang2019coupling}. In the weakly interacting regime, one can construct nonlinear solutions such as solitons from the harmonics of the topological edge states of the linear regime. Such situation was explored, for example, in the Kagome lattice \cite{Gulevich2017}. Recent works have predicted higher-order polariton topological insulators with corner states persisting in the non-linear regime \cite{Zhang2020}. 

\begin{figure}[t!]
\begin{center}
  \includegraphics[width=0.5\textwidth]{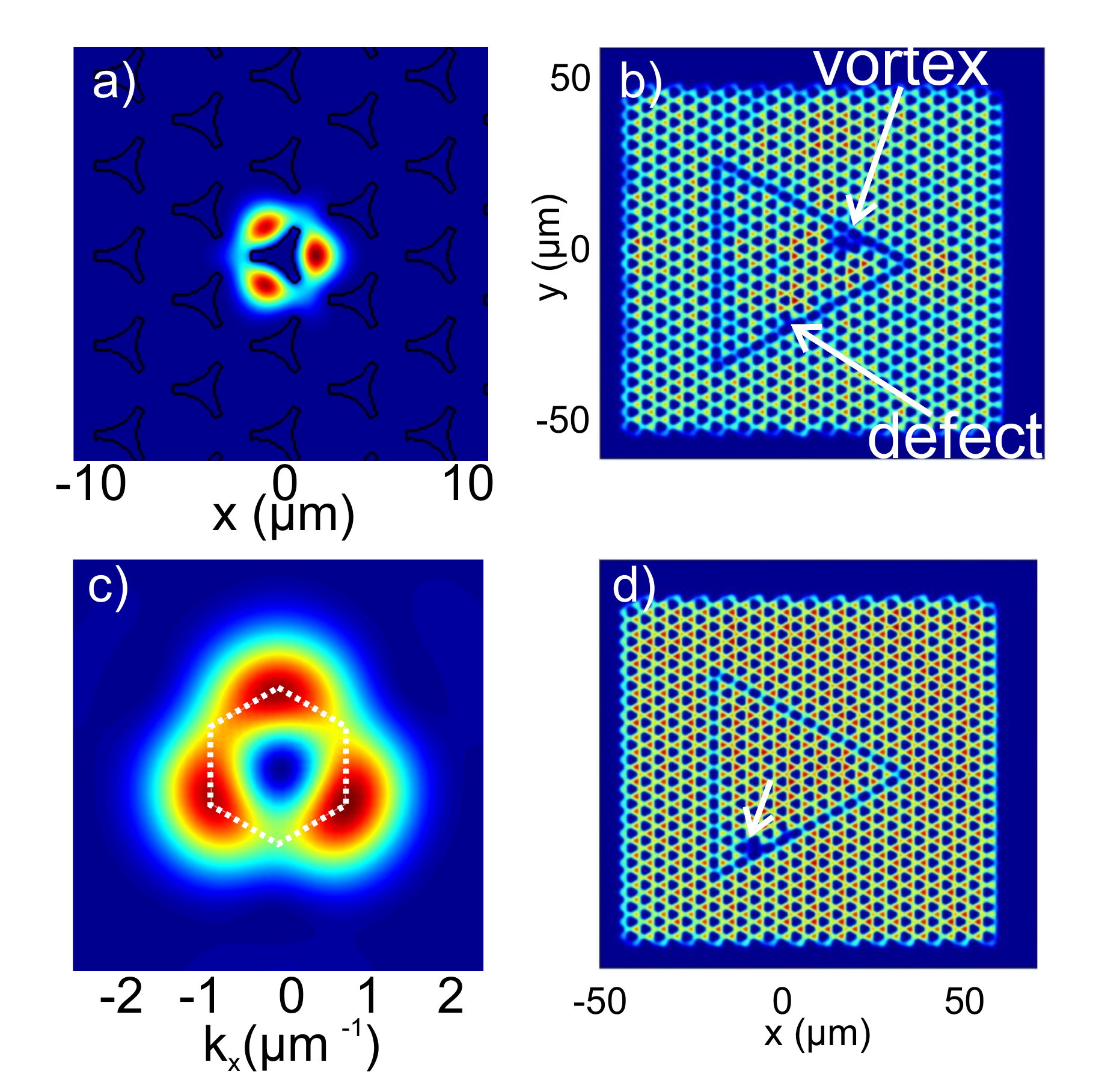}
  \caption{\label{fig:nlz2} Vortex core in real space (panel (a), potential shown by contour lines) and in reciprocal space (panel (c), localised at K points). The vortex is attached to the interface between two inverted staggered graphene lattices (b). The vortex (marked by an arrow) is not backscattered by defects (d). From Ref.~\cite{Bleu2018}.}
   \end{center}

 \end{figure}

The interplay of the topology of the lattice with the topology of the quantum fluid filling it has been shown to bring particularly interesting results~\cite{Bleu2018}. Indeed, the interface states that occur in the quantum valley Hall effect in a honeycomb lattice, which is simplest to implement since it does not require spin-orbit coupling nor time-reversal symmetry breaking by a magnetic field, are protected only by the symmetry of the lattice. Arbitrary disorder usually breaks this symmetry and, therefore, affects the interface states, coupling the opposite propagation directions by coupling the valleys \cite{Bleu2017}. This means that the chirality of the interface states can be broken by random disorder or isolated defects. On the other hand, vortices in the photon fluid are known to be protected by their own topological invariant, the winding number associated to the circulation around the core \cite{Pitaevskii2003}. If a polariton condensate is formed at the $\Gamma$ point of the polariton graphene, its vortices can be shown to be localised at the $K$ points and to exhibit a winding-valley coupling: the sign of the winding number determines the particular $K$-valley, as shown in Fig.~\ref{fig:nlz2}(a,c). This provides topological protection to the quantum valley Hall effect \cite{Bleu2018} (Fig.~\ref{fig:nlz2}(b,d))): the scattering of the vortex between the valleys is impossible because it would require changing its winding number, which is protected by the vortex topology.

\begin{figure*}[t!]
\begin{center}
  \includegraphics[width=1\textwidth]{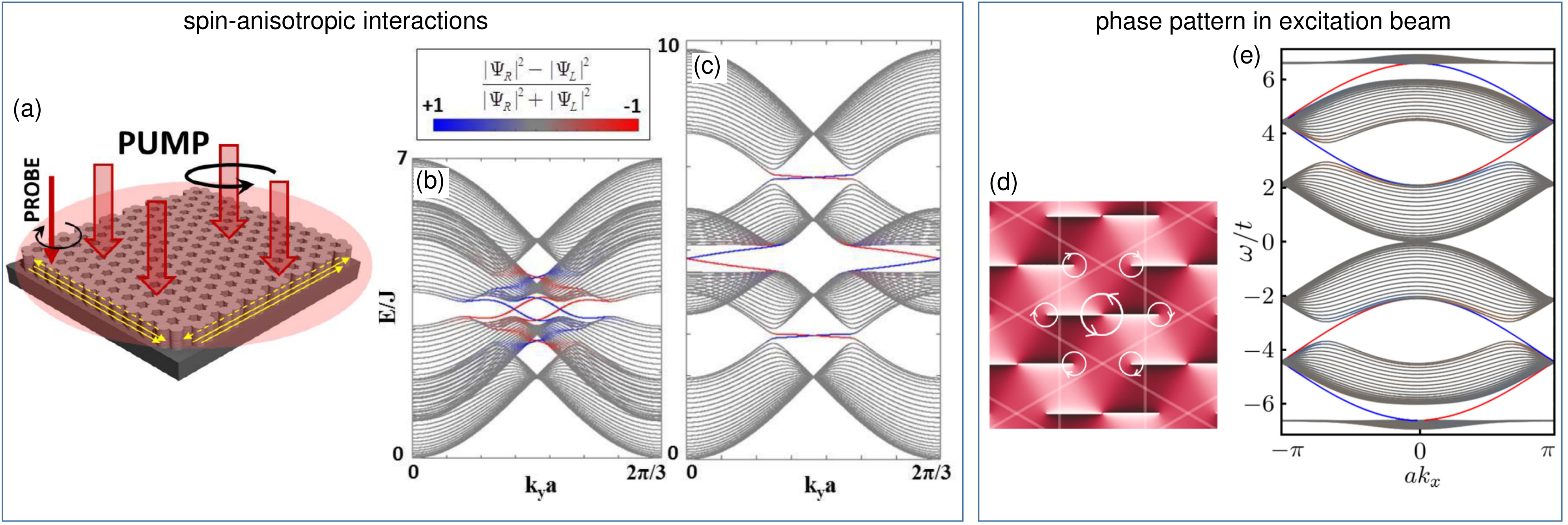}
  \caption{\label{fig:nonlinear} (a) Scheme of the excitation conditions to induce an optical Zeeman splitting in a polariton honeycomb lattice. The pump beam is circularly polarised at an energy in resonance with the lowest energy states of the lattice. Propagating edge states (yellow arrows) appear as a consequence of the topological gap in the spectrum of excitations, shown in (b) and (c) for moderate and high pump intensity, respectively. At moderate intensity, the bands around the central gap have a Chern number of $\pm2$, while at high densities the Chern number changes to $\mp1$ above a critical intensity at which the gap closes. From Ref.~\cite{Bleu2018b}. (d) Phase pattern of the resonant excitation beam in a Kagome lattice resulting in the emergence of topological gaps in the spectrum of excitations (e). Each arrow in (d) indicates a change of phase of $2\pi$. From Ref.~\cite{Bardyn2016}. Blue and red lines in (b), (c) and (e) label edge states located at the left and right edges respectively.}
   \end{center}
 \end{figure*}

\subsection*{Topology induced by nonlinearities: topological gap in the Bogoliubov spectrum}

In principle, nonlinear terms in the lattice Hamiltonian can break the underlying symmetries and modify the topological properties of the system. When an extended interacting polariton fluid is present in the lattice (i.e., a polariton condensate or a resonantly excited polariton fluid), the relevant spectrum of the system is the spectrum of excitations~\cite{Carusotto2004, Ciuti2005,Szymanska2006,Wouters2007}. In the context of ultracold atoms, it has been shown that the dispersion of the weak excitations on top of an underlying condensate inherits the topology of the original band dispersion \cite{Furukawa2015}, unless the interactions are themselves topologically nontrivial \cite{Sato2017}.

Interestingly, in the case of polariton fluids, it is possible to engineer emerging topology, that is, to reach non-trivial topological phases in the spectrum of excitations of an interacting fluid lying in an otherwise trivial lattice. As we will describe below, two ways to get there are the use of the polarisation dependent polariton-polariton interactions, and the fact that a spatial phase pattern can be imposed on the fluid using an external resonant laser. In both cases, time reversal symmetry is broken in the interacting fluid, and Chern insulating phases appear in the spectrum of excitations.

A rather unique property of polariton interactions is their polarisation anisotropic nature \cite{Renucci2005,Shelykh2006,Vladimirova2010}. Indeed, the interaction constant in the singlet configuration (opposite spins) is much weaker than that in the triplet configuration (polaritons with the same spin) because the virtual intermediate states of the involved exchange interaction for opposite spins are dark, purely excitonic states, which have much higher energies than the polaritonic states. Thus, if one creates polaritons with a given circular polarization, the energy of polaritons with this polarization strongly increases 
whereas the energy of the other component remains the same. In other words, a high density of circularly polarised polaritons results in a blueshift for polaritons with the same polarisation and almost no change in energy for polaritons of the opposite polarisation. Therefore, it is possible to efficiently break time-reversal symmetry by optical circular-polarised pumping, and induce an "optical Zeeman effect" similar to that observed with an external magnetic field.

Using this feature it is possible to implement a Chern insulating phase in a honeycomb lattice in which the effective Zeeman field is provided by the anisotropic interactions of an optically pumped polariton gas with a given circular polarisation. This situation was analyzed in a honeycomb lattice by Bleu and coworkers in Refs.~\cite{Bleu2016, Bleu2017x}.
Reference~\cite{Bleu2017x} considered the case of quasi-resonant pumping in which the excitation laser energy is quasi-resonant with a given polariton mode of the lattice. This situation is known to exhibit different possible behaviors, such as a spin-dependent bistability and multistability \cite{Gippius2007,Paraiso2011}. The dispersion of the density excitations of the polariton gas in these conditions strongly depends on the laser frequency \cite{Carusotto2004,Solnyshkov2008}. In order to guarantee the stability of the system (avoid bistable effects), the laser can be tuned below the bottom of the dispersion of a polariton honeycomb lattice. The circular-polarised pump is expected to induce an all-optical Zeeman splitting for the density excitations, breaking the time-reversal symmetry, and resulting in the opening of a topological gap at the Dirac crossings. The magnitude of the topological gap is directly controlled by the intensity of the laser, and it should make possible the observation of the topological transition with change of Chern numbers from $\pm 2$ to $\mp 1$ that takes place when increaing the Zeeman splitting. Indeed, the optical Zeeman splitting based on anistropic interactions is expected to be significantly larger than the Zeeman splitting induced by an external magnetic field. Across this topological transition, the topological edge modes change the propagation direction, which might be useful for applications \cite{Solnyshkov2018o}. A similar proposal to use polarised pumping to break the time-reversal symmetry, in the bistable regime, was made in another work \cite{Kartashov2007}.

A different regime is that of polariton condensation under nonresonant pumping~\cite{Kasprzak2006,Balili2007}. In this regime different scenarios have been shown to result in Chern-type topological bands based on the spin-anisotropic polariton interactions. The study of \cite{Bleu2016} deals with the case of thermal equilibrium of the condensate. In this case, at low external magnetic fields, the interaction induced Zeeman splitting compensates the external field, in a phenomenon known as the spin-Meissner effect \cite{Rubo2006,Gulevich2016b}. It turns out that the renormalization of the dispersion by the interactions depends on the wavevector: the spin-anisotropic interactions close the gap at the $K$ point faster than at the $\Gamma$ point, which allows observing a topological phase transition as a function of polariton density with the inversion of Chern numbers. If polariton condensation takes place in a kinetic regime (out of thermal equilibrium)~\cite{Levrat2010} spontaneous symmetry breaking of the spin of the condensate can take place~\cite{Ohadi2012} and result in spontaneous topological transitions~\cite{Sigurdsson2019}.

Other strategies have also been proposed to implement topological Chern phases that do not require the presence of a significant TE-TM SOC, which was required in the above mentioned examples. The idea is to inject polaritons in a lattice with a suitable phase pattern via a resonant excitation beam. In the nonlinear regime, the density excitations are subject to the phase gradients of the laser field resulting in the break up of time-reversal symmetry for the excitations. In a lattice, the Bogoliubov spectrum of excitations shows the opening of topological gaps at the band touchings and the appearance of superfluid chiral edge states~\cite{Bardyn2016,Sigurdsson2017}.

\section{Conclusion}

Polaritons offer virtually unlimited possibilities for wavefunction engineering, design of topological systems with promising applications, such as quantum opto-valleytronic systems. The natural presence of losses and the possibility of engineering almost any laser gain profiles, makes microcavity polaritons a very suitable experimental platform to explore non-Hermitian physics, and in particular their topological aspects. Polaritons also allow studying the interplay of the topology emerging in interacting quantum fluids and that of photonic lattices. Indeed, experimental evidence of topological effects induced by nonlinearities remains a largely unexplored area in which polaritons are one of the most promising platforms. Moreover, recent works have provided evidence that polariton nonlinearities can be significantly enhanced using indirect excitons~\cite{Togan2018, Rosenberg2018a}, which could be a promising route towards the quantum regime in order to investigate strongly correlated phases.

\section{Acknowledgements}
This work was supported by the H2020-FETFLAG project PhoQus (820392), the QUANTERA project Interpol (ANR-QUAN-0003-05), ERC CoG EmergenTopo (865151), the Marie Sklodowska-Curie individual fellowship ToPol, the French National Research Agency project Quantum Fluids of Light (ANR-16-CE30-0021), the Paris Ile-de-France Region in the framework of DIM SIRTEQ, the Labex NanoSaclay (ANR-10-LABX-0035) and CEMPI (ANR-11-LABX-0007), the French government through the Programme Investissement d’Avenir (I-SITE ULNE / ANR-16-IDEX-0004 ULNE) managed by the Agence Nationale de la Recherche and the CPER Photonics for Society P4S.
We also acknowledge the support of the projects EU "QUANTOPOL" (846353), of the ANR Labex GaNEXT (ANR-11-LABX-0014), and of the ANR program "Investissements d'Avenir" through the IDEX-ISITE initiative 16-IDEX-0001 (CAP 20-25). 
\noindent The authors declare no conflicts of interest.


\bibliography{alberto}

\end{document}